\documentclass[twocolumn]{svjour3}         
\smartqed  
\usepackage{graphicx}
 \usepackage{mathptmx}      
\usepackage{natbib}
\newcommand{\aap}{A\&A}

\newcommand{\apjl}{ApJL}
\newcommand{\gtsim}{{_>\atop{^\sim}}}
\begin{document}

\title{Molecular data and radiative transfer tools for ALMA
}


\author{Floris van der Tak         \and
        Michiel Hogerheijde 
}


\institute{F.\ van der Tak \at
              SRON Netherlands Institute for Space Research \\
              Landleven 12, 9747 AD Groningen, The Netherlands \\
              Tel.: +31-50-3638753\\
              Fax: +31-50-3634099\\
              \email{vdtak@sron.rug.nl}           
           \and
           M.\ Hogerheijde \at
           Leiden University Observatory \\
           P.O.\ Box 9513, 2300 RA Leiden, The Netherlands
}

\date{Received: 31 January 2007 / Accepted: ...}

\maketitle

\begin{abstract}

  This paper presents an overview of several modeling tools for analyzing
  molecular line observations at submillimeter wavelengths. These tools are
  already proving to be very valuable for the interpretation of data from
  current telescopes, and will be indispensable for data obtained with ALMA. The
  tools are: (1) the Leiden Atomic and Molecular DAtabase (LAMDA), a collection
  of spectroscopic data and collisional excitation rates; (2) RADEX, an on-line
  and off-line program to calculate non-LTE excitation and emission from a
  homogeneous medium, based on the escape probability approximation; (3) RATRAN,
  an accelerated Monte Carlo program to solve molecular excitation and radiative
  transfer in spherical and cylindrical symmetry. The paper presents examples of
  how to use these tools in conjunction with existing data reduction packages to
  quantitatively interpret submillimeter single-dish and interferometric observations. The
  described tools are publically available at
  \par \texttt{http://www.strw.leidenuniv.nl/$\sim$moldata}. 

  The paper concludes with a discussion of future needs in the fields of
  molecular data and radiative transfer.

\end{abstract}

\section{Introduction}
\label{intro}

Observations of spectral lines at radio, (sub)millimeter and infrared
wavelengths are powerful tools to investigate the physical and chemical
conditions in the dilute gas of astronomical sources where thermodynamic
equilibrium is a poor approximation \citep{genzel:crete1,black:korea}.
To extract astrophysical parameters from the data, the excitation and optical
depth of the lines need to be estimated, for which various methods may be
used, depending on the available observations \citep{vdishoeck:creteII,vdtak:catania}.

With the advent of ALMA, the need for efficient molecular line modeling tools
will become more pressing than ever. This paper presents several such tools
which we have developed and which we expect will become quite important for the
analysis of ALMA data.

\section{LAMDA}
\label{s:lamda}


The Leiden Atomic and Molecular Database (LAMDA) \citep{schoeier:lamda} contains
spectroscopic and collisional data for 24 species of astrophysical interest.
Most of these species are molecules, but data files for the fine structure lines
of C, O and C$^+$ are also provided. The data files contain energy levels,
statistical weights, Einstein $A-$coefficients and collisional rate
coefficients.  Available collisional data from quantum chemical calculations and
experiments have been extrapolated to higher energies (up to $E/k \sim$
1000\,K).  The format of the data files is such that they can be read directly
into RADEX and RATRAN.

\section{RADEX}
\label{s:radex}

\paragraph{Description} RADEX \citep{vdtak:radex} is a computer program to
calculate the intensities of atomic and molecular lines produced in a uniform
medium, based on statistical equilibrium calculations involving collisional and
radiative processes. The treatment includes radiation from background sources;
optical depth effects are calculated using the escape probability
approximation. The user has a choice of geometries: uniform sphere, expanding
sphere, or plane-parallel slab. The expanding sphere solution is analogous to
the well-known LVG method. 

\paragraph{Availability} The program is available both as an on-line calculator
for quick estimates and as an off-line package for detailed analysis of
multi-line datasets.

\paragraph{Typical Uses} (1) Multi-line observations of a region are available,
and an estimate is needed of the average temperature, density, and column
density.
(2) Based on estimates of the typical conditions in a source, estimates are made
of the expected line intensities, for example to prepare an observing proposal.

\begin{figure*}
  \includegraphics[width=0.75\textwidth]{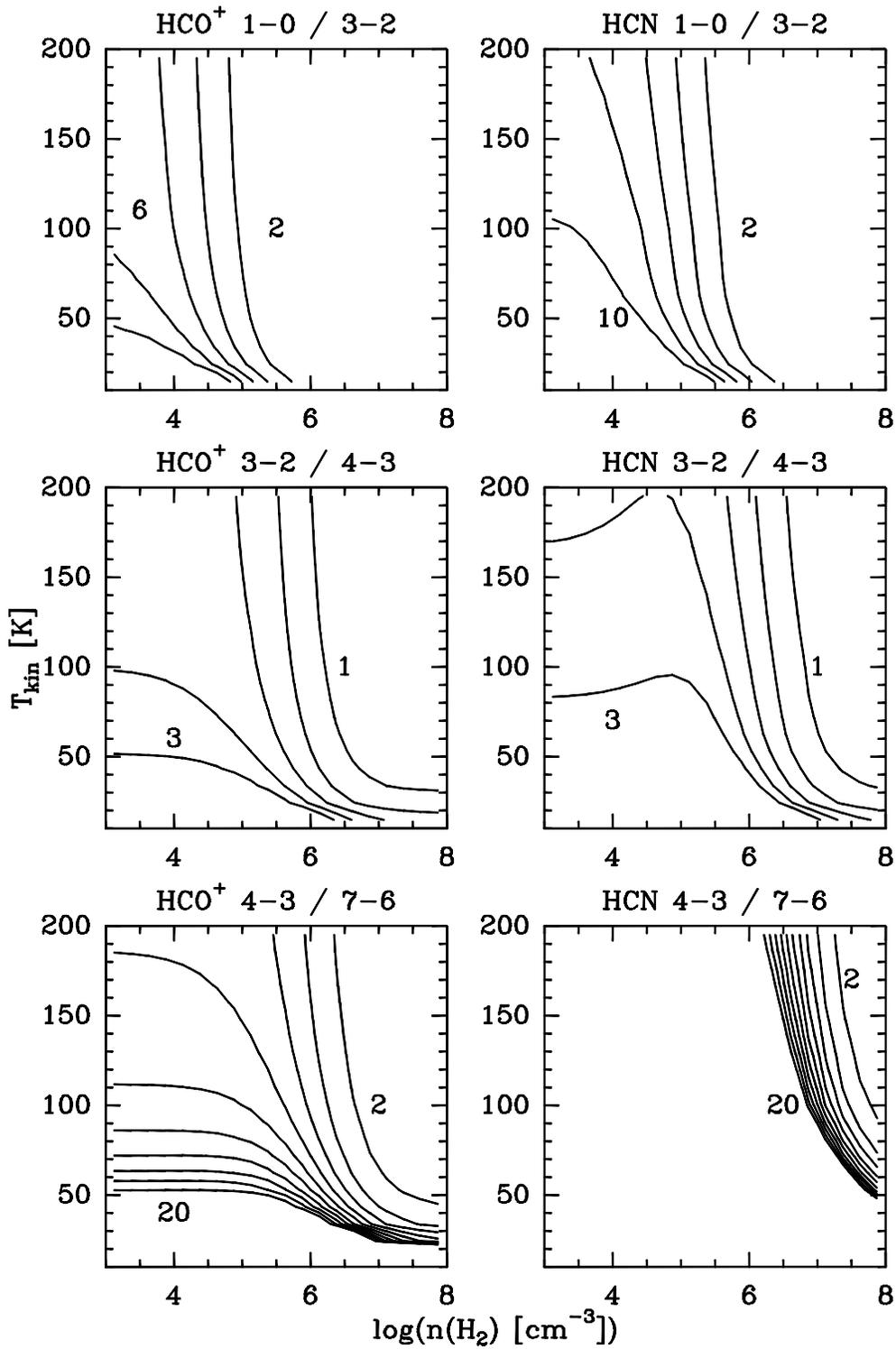}
  \caption{Line ratios of HCO$^+$ and HCN, calculated with RADEX in the optically thin
    limit as a function of kinetic temperature and H$_2$ volume density.
    Contours are spaced linearly and some contours are labeled for easy
    identification.}
\label{fig:ratios}       
\end{figure*}

\paragraph{Example} Ratios of submillimeter lines of HCN and HCO$^+$ trace
densities between 10$^3$ and 10$^8$ cm$^{-3}$ and temperatures between 20 and
200 K. By observing 3--4 different transitions, density and temperature can be
found (Fig.~\ref{fig:ratios}).

\section{RATRAN}
\label{s:ratran}

\paragraph{Description} RATRAN \citep{hogerheijde:montecarlo} is a computer
program to simulate non-LTE line radiative transfer in spherical or cylindrical
symmetry. The program uses the accelerated Monte Carlo method to represent the
radiation field, which offers both speed and reliability.

Within these geometries, the source model can have any density, temperature,
abundance, and velocity distribution. The program produces synthetic spectral
line cubes for a given source distance and orientation. Third-party software
such as Miriad or IDL readily produce higher-level output such as beam-convolved
line fluxes or even simulated interferometric visibilities.

\paragraph{Availability} The spherically symmetric version of the program is
available as an off-line package. The cylindrically symmetric version is
available on a collaborative basis. 

\paragraph{Typical Use} The user has obtained molecular line observations of an
object, and wishes to quantitatively test a particular model. With RATRAN, he
produces synthetic data and compares these to the observations by eye, or uses a
$\chi^2$ minimization to arrive at an optimum fit.

\paragraph{Example} \citet{brinch:l1489} compare observations of the Young
Stellar Object L~1489~IRS to a model with infalling and rotational motions.
The data set consists of millimeter interferometric observations, near-infrared
scattered light imaging, and CO ro-vibrational lines. Comparison of data and
model uses a Voronoi optimisation algorithm.

\section{Future molecular data needs}
\label{s:future}

Of the 134 molecules which are known in space, collisional rate coefficients
(CRCs) exist only for twenty-four (\S~\ref{s:lamda}). 
Accurate CRCs are important for reliable estimates of molecular
column densities, and essential to estimate other parameters such as
kinetic temperature and volume density.
Recently several important molecules for which only inaccurate CRCs
 were known, based on outdated, low-quality potential
energy surfaces, have been re-calculated, such as CS, HC$_3$N, and
CH$_3$OH \citep{dubernet:collrates}.

Calculation of rate coefficients is a highly specialized and time
consuming effort. New results appear in the literature at a rate of a
few molecules per year, which is at or below the detection rate of new
interstellar molecules. The point where CRCs are known for all
interstellar molecules may thus never be reached.
In order to make the best possible use of scarce human and
computational resources, we have selected those molecules
for which the calculation of new CRCs would have the most
astrophysical impact.


The main criterion to assign high priority to a molecule is its importance in
constraining astrophysical or astrochemical parameters beyond its own abundance.
Calculation of CRCs is most useful for species which uniquely trace certain
physics in a wide variety of astrophysical objects.

As a second criterion, the possibility to use old results or results from other
molecules is considered.
Although direct calculations are always preferred over approximate
scalings, the impact of new calculations is somewhat less if scaling
may be expected to yield fairly accurate approximations.
%
The most common examples of such scalings are isotopologues, where CRC
scaling will only work if the symmetry of the molecule is unchanged
(e.g., HCO$^+$$\to$DCO$^+$ will work but H$_2$CO$\to$HDCO will not);
isomers, where scaling only works if the species have similar dipole
moments (e.g., HCN$\to$HNC works better than HCO$^+$$\to$HOC$^+$);
and O$\to$S substitutions (e.g., H$_2$CO$\to$H$_2$CS).
Unlike in molecular physics, accuracy to order of magnitude is often
enough in astrophysics, and relative rates are more important than
absolute values.

The astrophysically most important molecules without any known rate
coefficients and without structurally similar species with known rates are:

\begin{enumerate}
\item \textbf{CN} is a key tracer of energetic radiation in protoplanetary
  disks, Galactic PDRs and extragalactic systems especially when compared with
  HCN \citep{meijerink+spaans}. Rate coefficients for electron collisions are
  known \citep{black:cn}, but no values for H$_2$ or He exist. Scaling from CO or
  CS fails because CN has a $^2\Sigma$ ground state. Note that calculated rates
  may be scalable to CO$^+$, another tracer of high-energy radiation, recently
  detected in the active galaxy M~82 \citep{fuente:m82}.

\item \textbf{H$_2$D$^+$ and D$_2$H$^+$} are unique chemical and kinematic tracers of
  the centers of pre-stellar cores, where all CNO-bearing species are frozen out
  on dust grains \citep{caselli:h2d+,vdtak:l1544,vastel:d2h+}. The asymmetric
  light-weight structure prevents scaling from other molecules.

\item \textbf{NO} is another readily observed important PDR and XDR tracer.
  Information from other $^2\Pi$ molecules such as OH cannot be scaled
  because of the different Hund case. Experimental cross sections have
  been measured by the group of Milliard Alexander in Maryland, and
  may be used as a starting point. The rates may be further scaled to
  NS, a possible shock tracer \citep{hatchell+viti}.

\item \textbf{H$_3$O$^+$} is important as tracer of H$_2$O and of the local
  ionization rate \citep{vdtak:h3o+}. So far, scaled NH$_3$ rates have been used
  \citep{phillips:h3o+} which are at best correct to order of magnitude. 
\end{enumerate}


Finally it would be useful to develop approximations to derive
rate coefficients for large molecules ($\gtsim$7 atoms) for which
detailed calculations would take prohibitive amounts of time.

\begin{acknowledgements}
  The research of the authors is supported by the Netherlands Organization for
  Scientific Research (NWO) and the Netherlands Research School for Astronomy
  (NOVA).
\end{acknowledgements}

\bibliographystyle{aa}      
\bibliography{vdtak}   

\end{document}